\documentclass[%
 reprint,
superscriptaddress,
 amsmath,amssymb,
 aps,
floatfix,
]{revtex4-2}

\usepackage{graphicx}% Include figure files
\usepackage{dcolumn}% Align table columns on decimal point
\usepackage{bm}% bold math
\usepackage{color}
\usepackage{here}
\usepackage{ulem}

\begin{document}

\preprint{APS/123-QED}

\title{Anisotropic response of spin susceptibility in the superconducting state of UTe${}_2$ probed with $^{125}$Te-NMR measurement}% Force line breaks with \\

\author{Genki Nakamine}
\affiliation{%
{\it Department of Physics, Kyoto University, Kyoto 606-8502, Japan}
}%
\author{Katsuki Kinjo}
\affiliation{%
{\it Department of Physics, Kyoto University, Kyoto 606-8502, Japan}
}%
\author{Shunsaku Kitagawa}
\affiliation{%
{\it Department of Physics, Kyoto University, Kyoto 606-8502, Japan}
}%
\author{Kenji Ishida}
\affiliation{%
{\it Department of Physics, Kyoto University, Kyoto 606-8502, Japan}
}%
\author{Yo Tokunaga}%
\affiliation{%
{\it ASRC, Japan Atomic Energy Agency, Tokai, Ibaraki 319-1195, Japan}
}%
\author{Hironori Sakai}
\affiliation{%
{\it ASRC, Japan Atomic Energy Agency, Tokai, Ibaraki 319-1195, Japan}
}%
\author{Shinsaku Kambe}
\affiliation{%
{\it ASRC, Japan Atomic Energy Agency, Tokai, Ibaraki 319-1195, Japan}
}%

\author{Ai Nakamura}
\affiliation{%
{\it IMR, Tohoku University, Oarai, Ibaraki 311-1313, Japan}
}%
\author{Yusei Shimizu}
\affiliation{%
{\it IMR, Tohoku University, Oarai, Ibaraki 311-1313, Japan}
}%
\author{Yoshiya Homma}
\affiliation{%
{\it IMR, Tohoku University, Oarai, Ibaraki 311-1313, Japan}
}%
\author{Dexin Li}
\affiliation{%
{\it IMR, Tohoku University, Oarai, Ibaraki 311-1313, Japan}
}%
\author{Fuminori Honda}
\affiliation{%
{\it IMR, Tohoku University, Oarai, Ibaraki 311-1313, Japan}
}%
\author{Dai Aoki}
\affiliation{%
{\it IMR, Tohoku University, Oarai, Ibaraki 311-1313, Japan}
}%
\affiliation{%
{\it University Grenoble, CEA, IRIG-PHERIQS, F-38000 Grenoble, France}
}%

\date{\today}% It is always \today, today,

\begin{abstract}
To investigate spin susceptibility in a superconducting (SC) state, we measured the $^{125}$Te-nuclear magnetic resonance (NMR) Knight shifts at magnetic fields ($H$) up to 6.5 T along the $b$ and $c$ axes of single-crystal UTe$_2$, a promising candidate for a spin-triplet superconductor. 
In the SC state, the Knight shifts along the $b$ and $c$ axes ($K_b$ and $K_c$, respectively) decreased slightly and the decrease in $K_b$ was almost constant up to 6.5 T. 
The reduction in $K_c$ decreased with increasing $H$, and $K_c$ was unchanged through the SC transition temperature at 5.5 T, excluding the possibility of spin-singlet pairing. 
Our results indicate that spin susceptibilities along the $b$ and $c$ axes slightly decrease in the SC state in low $H$, and the $H$ response of SC spin susceptibility is anisotropic in the $bc$ plane.        
We discuss the possible $\mbox{\boldmath $d$}$-vector state within the spin-triplet scenario and suggest that the dominant $\mbox{\boldmath $d$}$-vector component for the case of $H \parallel b$ changes above 13 T, where $T_{\rm c}$ increases with increasing $H$.

\end{abstract}

\maketitle
Newly discovered UTe$_2$ with an orthorhombic structure has attracted increasing attention owing to its unusual superconducting (SC) properties\cite{RanScience2019, AokiJPSJ2019}.    
It can exhibit superconductivity at magnetic fields ($H$) that are much higher than the Pauli-limiting field, which can be calculated from the SC transition temperature $T_{\rm c}$, and $H$-boosted superconductivity appears when $H \parallel b$\cite{RanScience2019, KnebeJPSJ2019}.
A novel re-entrant superconductivity was also reported for a case when $H$ was applied at a certain angle to the $bc$ plane\cite{RanNaturePhys2019}.   
Although no ferromagnetic ordering was observed near the SC phase in UTe$_2$, such robustness of superconductivity against $H$ seems to be a common feature in uranium-based ferromagnetic and nearly ferromagnetic superconductors\cite{AokiJPSJRev2019}.
In addition, pressure measurements revealed multiple SC phases\cite{CommPhysBraithwaite2019, RanPhysRevB2020, thomasArXiv2020, AokiJPSJ2020Pressure}, which can be tuned by $H$\cite{LinArXiv2020, AokiJPSJ2020Pressure}, and a presumably magnetic ordered phase above 1.8 GPa\cite{thomasArXiv2020, KnebelJPSJ2020, AokiJPSJ2020Pressure}.    
Furthermore, spontaneously broken time-reversal symmetry, as revealed from the Kerr measurement,\cite{IanArXiv2020} and the presence of the chiral Majorana edge and surface state, as observed using scanning tunneling spectroscopy,\cite{JiaoNature2020} were suggested.
UTe$_2$ is like a ``toy box'' for condensed matter physicists as it storages most of fascinating topics at present.

In a previous study\cite{NakamineJPSJ2019}, we reported that UTe$_2$ is classified as an unconventional superconductor because of the absence of a coherence peak just below $T_{\rm c}$; moreover, the small decrease in the Knight shift when $H \parallel b$, which is much smaller than the expected decrease in a spin-singlet superconductor, is favorable for a spin-triplet scenario, because the large spin component remains in the SC state.
Similar Knight-shift behaviors were also observed in UPt$_3$\cite{TouPRL96, TouPRL98} and UCoGe\cite{HattoriPRB2013, ManagoPRB2019}.
To ensure spin triplet superconductivity, it is important to identify the spin state of the SC state.
For this purpose, UTe$_2$ is one of the most suitable superconductors to study physical properties of spin-triplet pairing because superconductivity emerges from the paramagnetic state without the effect from the ordered state.  
    
Spin-triplet superconductivity possesses spin degrees of freedom, which can be expressed as the $\mbox{\boldmath $d$}$ vectors perpendicular to the spin components of the spin-triplet pairing\cite{LegettRevModPhys1975}.
If the $\mbox{\boldmath $d$}$ vector is fixed to one of the crystalline axes by some mechanism, the spin part of the Knight shift decreases in the SC state when the magnetic field  $\mbox{\boldmath $H$}$ is parallel to the $\mbox{\boldmath $d$}$ vector ($\mbox{\boldmath $H$} \parallel  \mbox{\boldmath $d$}$); however, it remains unchanged in the case of  $\mbox{\boldmath $H$} \perp \mbox{\boldmath $d$}$.
The $\mbox{\boldmath $d$}$-vector component can be derived from the measurements of the Knight shift along each crystalline axis.
In the previous study, we performed NMR measurements in $H \parallel b$ on the single-crystal sample prepared using a natural tellurium source with 7.1\% $^{125}$Te, which is a nuclear magnetic resonance (NMR)-active isotope. 
To identify the spin state thoroughly, NMR measurements when $H \parallel c$ and $H \parallel a$ are needed.
However, the NMR intensity when $H \parallel c$ was not sufficient to obtain reliable results as the NMR spectrum in $H \parallel c$ is broader than that in  $H \parallel b$. 
This is because of the larger spin susceptibility in the $c$ axis than that in the $b$ axis at low temperatures.
Thus, in the present study, we used a $^{125}$Te-enriched single-crystal sample that contained 99.9\% of $^{125}$Te.  
This ensured more reliable NMR measurements because the signal intensity of the present sample was roughly one order of magnitude larger than that of the previous sample. 

Herein, we report the $H$ dependence of the Knight shift in the SC state along the $b$ and $c$ axes, since the NMR spectrum when $H \parallel a$ could not be observed below 20 K due to the divergence of the nuclear spin-spin relaxation rate $1/T_2$\cite{TokunagaJPSJ2019}. 
The findings by the present measurements support the existence of spin-triplet pairing and reveal that the $\mbox{\boldmath $d$}$-vector includes the $\hat{b}$ and $\hat{c}$ components that show an anisotropic response against $H$ in the $bc$ plane.  

%%%%%%%%%%%%%%%%%%%%%%%%%%%%%% Fig 1 %%%%%%%%%%%%%%%%%%%%%%%%%%%%%%%%%%%%%%
\begin{figure}[H]
\includegraphics[width=80mm]{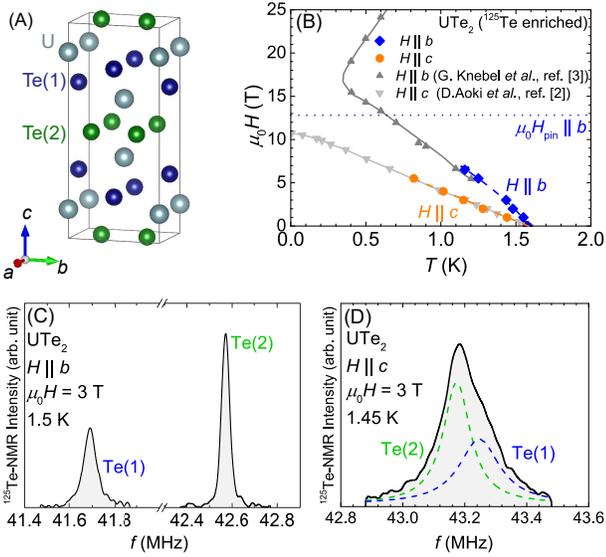}% Here is how to import EPS art
\caption{\label{fig:fig1}(A) Image of the crystal structure of UTe${}_2$ prepared with the program VESTA\cite{VESTA}.
(B) Upper critical field $H_{\rm {c2}}$ for a magnetic field applied along the $b$ (diamonds) and $c$ (circles) axes. 
The solid circles denote $T_{\rm c}$ of the $^{125}$Te-enriched sample, which is consistent with previous results\cite{AokiJPSJ2019, KnebeJPSJ2019}.
The dotted line indicates the $\mbox{\boldmath $d$}$-vector pinning field $H_{\rm pin}$ for $H \parallel b$ estimated from the decrease in $K_b$ and $A_{{\rm hf}, b}$. 
(C, D) $^{125}$Te NMR spectra measured when (C)$H \parallel b$ and (D)$H \parallel c$, respectively. 
The dotted peaks show signals from the Te(1) and Te(2) sites, which are simulated with the resonance frequencies estimated from the angle dependence of the spectrum shown in the supplemental material\cite{supplemental}. }
\end{figure}
%%%%%%%%%%%%%%%%%%%%%%%%%%%%%%%%%%%%%%%%%%%%%%%%%%%%%%%%%%%%%%%%%%%%%%%%

Single-crystal UTe$_2$ was grown using the chemical transport method with iodine as the transport agent.
Natural uranium and $^{125}$ Te-enriched metals were used as starting materials for the present sample.  
The $^{125}$Te ($I$ = 1/2, gyromagnetic ratio $^{125}\gamma_n$/2$\pi$ = 13.454 MHz/T)-NMR measurements were performed on a single crystal of size  3.5$\times$0.7$\times$1.4 mm$^3$.
The frequency-swept NMR spectrum was obtained using the Fourier transform of a spin-echo signal observed after the spin-echo RF pulse sequence with a 3 kHz step in a fixed magnetic field.
The magnetic field was calibrated using the $^{65}$Cu ($^{65}\gamma$/2$\pi$ = 12.089 MHz/T) NMR signal from the NMR coil. 
The NMR spectrum in the SC state was recorded with the field-cooling process. 
We used the split SC magnet, which generated a horizontal field, and combined it with a single-axis rotator to apply a magnetic field exactly parallel to the $b$ or $c$ axis; the $a$-axis was the rotation axis. 
Low-temperature NMR measurements down to 70 mK were performed using a $^3$He-$^4$He dilution refrigerator, and the single-crystalline sample was immersed into the mixture. 
The AC susceptibility was measured by recording the resonance frequency of the NMR-tank circuit during cooling to determine the $T_{\rm c}$ value of the sample under $H$.

UTe$_2$ exhibits two crystallographically inequivalent Te sites, 4$j$ and 4$h$, with point symmetries $mm$2 and $m$2$m$, respectively. 
We denote these sites as Te(1) and Te(2), respectively, as shown in Fig.~1 (A).
The $T_{\rm c}$ of the present sample determined by the AC susceptibility measurement was consistent with the previous results, as shown in Fig.~1 (B)\cite{AokiJPSJ2019, KnebeJPSJ2019}.
Figures 1 (C) and (D) show the frequency-swept NMR spectrum measured at 3 T when $H \parallel b$ (C) and $H \parallel c$ (D), respectively.
The Te(1) and Te(2) signals were distinct when $H \parallel b$ but overlapped when $H \parallel c$.
Thus, the Te(2) signal was recorded when $H \parallel b$, and the broad peak consisting of the Te(1) and Te(2) signals was recorded when $H \parallel c$ for the Knight shift and linewidth measurements.
The Knight shift at the Te(2) site was determined from the spectral-peak frequency because the contribution of the Te(2) site was dominant even in the broad spectrum when $H \parallel c$.

%%%%%%%%%%%%%%%%%%%%%%%%%%%%%% Fig 2  %%%%%%%%%%%%%%%%%%%%%%%%%%%%%%%%%%%%%%%%%%%
\begin{figure}[H] 
\includegraphics[width=85mm]{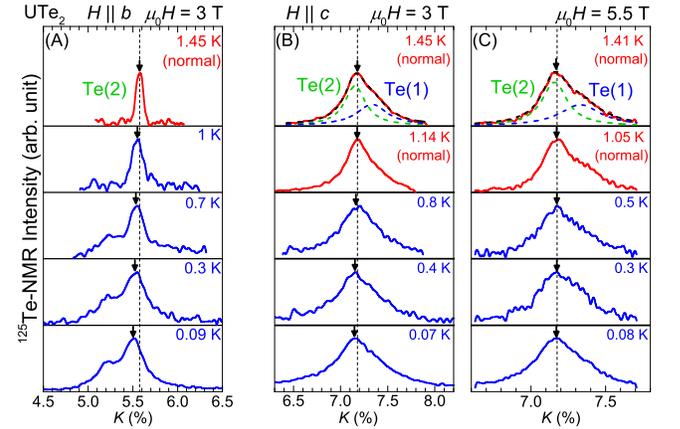}% Here is how to import EPS art
\caption{\label{fig:fig2}NMR spectra at several temperatures at 3 T along the $b$ (A) and $c$ (B) axes. NMR spectra at 5.5 T along the $c$ axis (C). 
The dotted line in each figure represents the normal-state peak position. }
\end{figure}
%%%%%%%%%%%%%%%%%%%%%%%%%%%%%%%%%%%%%%%%%%%%%%%%%%%%%%%%%%%%%%%%%%%%%%%%%%%
Figures \ref{fig:fig2} (A) and (B) indicate the NMR spectra measured at several temperatures at 3 T in $H \parallel b$ (A) and $H \parallel c$ (B), respectively, which are shown against $K \equiv$ ($f - f_0$) / $f_0$. 
Here, $f$ is the NMR frequency, and $f_0$ is the reference frequency determined as $f_0 = (\gamma_n / 2\pi) H$. 
Both spectra widen from just below $T_{\rm c}$ as shown in the supplemental material\cite{supplemental}, which ensures that the NMR spectrum in the SC state was measured.
In addition, shoulder peaks appear in the $H \parallel b$ spectrum below 0.9 K.
Because this may suggest the presence of multiple SC phases in $H$, the origin of the shoulder peaks must be thoroughly investigated.
In this study, we focus on the temperature variation of the central peak, and the details of the shoulder peaks will be summarized in a separate study.         
Notably, the main peak of the $H \parallel b$ spectrum, depicted with an arrow, shifts in the SC state, which is visible as a shift from the dotted line showing the normal-state Knight shift.   
In contrast, the peak of the $H \parallel c$ spectrum, shown by an arrow, slightly shifts in the SC state at 3 T [Fig.~2 (B)]; however, the shift of the peak could not be recognized in the $H \parallel c$ spectrum measured at 5.5 T, although the broadening of the spectrum was observed in the SC state [Fig.~2 (C)].  
Subsequently, $K_i$ ($i$ = $b$ and $c$) is determined by the peak frequencies shown by the arrows.

%%%%%%%%%%%%%%%%%%%%%%%%%%%%% Fig 3 %%%%%%%%%%%%%%%%%%%%%%%%%%%%%%%%%%%%%%%%%%%%%%%%%%%%%%%%%%%%%%
\begin{figure}[H]
\includegraphics[width=85mm]{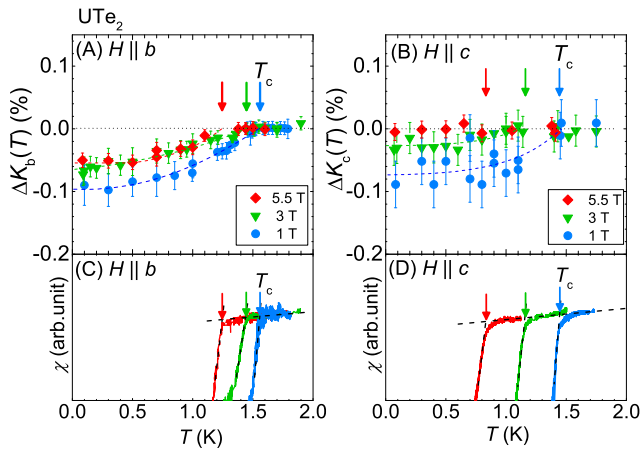}% Here is how to import EPS art
\caption{\label{fig:fig3} Temperature dependence of $\Delta K_b$ (A) and $\Delta K_c$ (B) measured at 1, 3, and 5.5 T (see in the text).
Dashed lines are added as a guide to the eye. 
Temperature dependence of the AC susceptibility $\chi$ at the same magnetic fields along the $b$ (C) and $c$ (D) axes. 
The errors of the NMR Knight shift are determined from the resolution of FT signals.}
\end{figure}
%%%%%%%%%%%%%%%%%%%%%%%%%%%%%%%%%%%%%%%%%%%%%%%%%%%%%%%%%%%%%%%%%%%%%%%%%%%%%%%%%%%%%%%%%%%%%%%
We investigated the temperature variation of $K_b$ and $K_c$ measured at several magnetic fields below 5.5 T.
Figures \ref{fig:fig3} (A) and (B) represent the temperature variation of $\Delta K_b(T)$ (A) and $\Delta K_c(T)$ (B) at $\mu_0 H$ = 1, 3, and 5.5 T, which are compared with the temperature dependence of the AC susceptibility.
Here, $\Delta K_i (T)$ ($i$ = $b$ and $c$) is defined as $\Delta K_i (T) \equiv K_i (T) - K_{n, i}$ with the normal-state Knight shift denoted as $K_{n, i}$. 
The decrease in $\Delta K_b(T)$ observed at 1 T is $\sim$ 0.1\%, which is consistent with the previous result\cite{NakamineJPSJ2019} and the recent theoretical calculation based on the spin-triplet state\cite{HiranumaArXiv2020}.
Further reductions at a low temperature, related to the multi-gap properties, were not observed in the present sample as well as in the previous sample.  
Notably, the decrease observed at 1 T is almost the same in $\Delta K_b$ and $\Delta K_c$.
The decreases in $\Delta K_b(T)$ at 3 and 5.5 T in the SC state, which are slightly smaller than that at 1 T, appear to be inconsistent with the previous result\cite{NakamineJPSJ2019}. 
This could be due to the fitting of the whole spectrum for the estimation of the Knight shift in the previous measurement on the $^{125}$Te natural-abundant sample.
The decrease in $\Delta K_b$ in the SC state seems to be almost $H$-independent above 3 T.  
In contrast, $\Delta K_c$ at the lowest temperatures became zero with an increase in $H$, as shown in Fig.~\ref{fig:fig3}~(B).

%%%%%%%%%%%%%%%%%%%%%%%%%%% Fig4 %%%%%%%%%%%%%%%%%%%%%%%%%%%%%%%%%%%%%%%%%%%%%%%%%%%%%%%%%%%%%%%%%%%
\begin{figure}
\includegraphics[width=70mm]{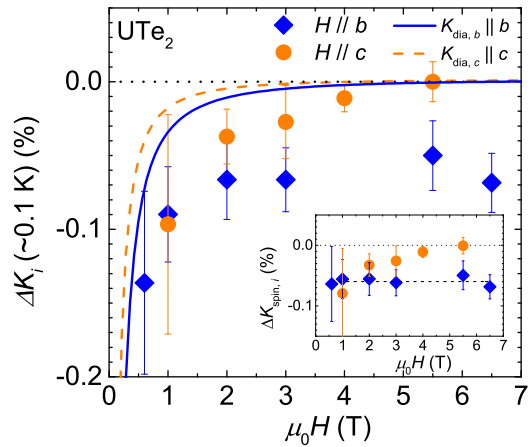}% Here is how to import EPS art
\caption{\label{fig:fig4}
The magnetic field dependence of the decrease in $\Delta K_b$ (diamonds) and $\Delta K_c$ (circles).
The solid and dashed lines represent the calculated SC diamagnetic shielding effects $\Delta K _{{\rm dia}, b}$ and $\Delta K _{{\rm dia}, c}$, respectively.
The inset shows the $H$ dependence of the decrease in $K_{{\rm spin}, i}$ ($\Delta K_{{\rm spin}, i}$) estimated by subtracting $\Delta K_{{\rm dia}, i}$ from $\Delta K_i$. }
\end{figure}
%%%%%%%%%%%%%%%%%%%%%%%%%%%%%%%%%%%%%%%%%%%%%%%%%%%%%%%%%%%%%%%%%%%%%%%%%%%%%%%%%%%%%%%%%%%%%%%%%
To observe the $H$ dependence of the $K_i$ decrease in the SC state more quantitatively, $\Delta K_i$ at the lowest temperature ($\sim 0.1$ K), $\Delta K_i(\sim 0.1 {\rm K})$  is plotted against $H$ in Fig.~\ref{fig:fig4}.
In general, $\Delta K_i$ is attributed to two contributions: a decrease in the spin part of $K_i$ ($\Delta K _{{\rm spin,} i}$) in the SC state and the SC diamagnetic shielding effect $\Delta K _{{\rm dia}, i}$, which is expressed as \cite{deGennesSC} 
\begin{equation} 
\Delta K_{{\rm dia}, i} = \frac{H_{c1, i}}{H} \frac{\ln(\frac{\beta d}{\sqrt{e} \xi})}{\ln{\kappa}}.
\end{equation}
Here, $\xi$ is the GL coherence length; $\beta$ is a factor depending on the vortex structure and is 0.38 for the triangular vortex lattice, and $d$ is the distance between vortices and is calculated using the relation $\phi_0 = \frac{\sqrt{3}}{2} d^2(\mu_0H).$
We estimate the $H$ dependence of $\Delta K_{{\rm dia}, i}$ ($i$ = $b$ and $c$) as follows: 
In the estimation of  $\Delta K_{{\rm dia}, c}$, we use the SC critical field ($H_c$) of UTe$_2$, which is reported to be 49 mT\cite{PaulsenArXiv2020}, and the upper critical field ($H_{\rm {c2}}$) along the $c$ axis, which is $\sim 11$ T, as shown in Fig.~1 (B).
This gives $\mu_0 H_{{c1}, c}$ =  1.2 mT, $\xi$ = 5.38 nm, and $\kappa$ = 159, and the $H$ dependence of $\Delta K_{{\rm dia}, c}$ is shown by the dashed curve in Fig.~\ref{fig:fig4}.
For the estimation of $\Delta K_{{\rm dia}, b}$, we refer to $\mu_0 H_{c1}$ = 2 mT\cite{PaulsenArXiv2020}, and $H_{\rm {c2}}$ along the $b$ axis is assumed to be $\sim$ 17 T from the extrapolation of $H$ dependence of $T_{\rm c}$ in the low-$H$ region.
The $H$ dependence of $\Delta K_{{\rm dia}, b}$ is indicated by the solid curve in Fig.~\ref{fig:fig4}. 
As highlighted by Paulsen {\it et al.}, the $H_{\rm c1}$ value along the $b$ axis is unexpectedly large and is two times the estimated value from $H_c$\cite{PaulsenArXiv2020}.
Even if such an unexpectedly large $H_{c1}$ is adopted for $H \parallel b$, the observed $H$ dependence of the $K_b$ decrease cannot be explained solely based on the $H$ dependence of $\Delta K_{{\rm dia}, b}$; however, it indicates that $K_{\rm spin}$ along the $b$ and $c$ axes decreases in the SC state, and the decrease in $K_{\rm spin}$ shows an anisotropic response against applied $H$, as shown in the inset of Fig.~\ref{fig:fig4}.    
This suggests that the decrease in the spin susceptibility is maintained at least up to 6.5 T when $H \parallel b$; in contrast, it gradually becomes small when $H \parallel c$, and the spin susceptibility remains unchanged with temperature at 5.5 T.                

%%%%%%%%%%%%%%%%%%%%%%%%%%%%%%%%%%%%% table %%%%%%%%%%%%%%%%%%%%%%%%%%%%%%%%%%%%%%%%%%%%%%%%%%%%%%
\begin{table}
\centering
\caption{\label{tab:op}Classification of odd-parity SC order parameters for point groups with $D_{2h}$.\cite{IshizukaPRL2019}}
\begin{tabular}{cc}\hline \hline
Irreducible representation \hspace{1cm}  & Basis function  \\ \hline
$A_u$    & $k_a$$\hat{a}$, $k_b$$\hat{b}$, $k_c$$\hat{c}$   \\
$B_{1u}$ & $k_b$$\hat{a}$, $k_a$$\hat{b}$ \\
$B_{2u}$ & $k_a$$\hat{c}$, $k_c$$\hat{a}$ \\
$B_{3u}$ & $k_c$$\hat{b}$, $k_b$$\hat{c}$ \\ \hline \hline
\end{tabular}
\end{table}

%%%%%%%%%%%%%%%%%%%%%%%%%%%%%%%%%%%%%%%%%%%%%%%%%%%%%%%%%%%%%%%%%%%%%%%%%%%%%%%%%%%%%%%%%%%%%%%

We discuss the plausible SC state based on the present experimental results.
As mentioned in the introduction, the SC pairing in UTe$_2$, at least at ambient pressure, has been considered to be a spin-triplet state.
$K_c$ remains unchanged with temperature in the SC state at 5.5 T; this excludes the possibility of spin-singlet pairing even though vortex and/or the $H$-induced quasiparticle contribution is considered. This is because the decrease in the spin susceptibility should be observed in all directions for spin-singlet pairing.
This further supports the spin-triplet scenario.
With regard to the spin-triplet superconductivity with odd parity, the possible SC order parameters were highlighted from an ordinary classification theory with $D_{2h}$ point group symmetry, which are shown in TABLE \ref{tab:op}\cite{IshizukaPRL2019}.
Since no indication of non-unitary states was observed just below $T_{\rm c}$ in our measurements, non-unitary states are not considered below.
The spin susceptibility decreases with a similar magnitude for $H \parallel b$ and $H \parallel c$ in low $H$; this suggests that the $\mbox{\boldmath $d$}$-vector contains finite $\hat{b}$ and  $\hat{c}$ components in low $H$.
If the strong Ising anisotropy along the $a$ axis in the normal-state spin susceptibility is considered, the $\mbox{\boldmath $d$}$ vector is expected to be perpendicular to the $a$ axis; thus, the $B_{3u}$ state is a promising candidate.
This seems to be consistent with the point-node gap suggested by angle-resolved specific-heat measurements\cite{KittakaPRR2020}.
However, because the relationship between the normal-state spin and the $\mbox{\boldmath $d$}$-vector anisotropy is not clear at present, the $A_u$ state might also be possible.
In this case, the component of each basis is highly anisotropic in the $\mbox{\boldmath $k$}$ space. 
To distinguish the two possibilities, the Knight shift measurement along the $a$ axis would provide important information regarding the $\hat{a}$ component, if it can be measured.

Furthermore, the anisotropic response of the SC spin susceptibility in the $bc$ plane against the applied $H$ suggests that the $\hat{c}$ component in the $\mbox{\boldmath $d$}$-vector becomes small when $H \parallel c$ and is nearly zero at 5.5 T; moreover, the $\hat{b}$ component is robust for $H \parallel b$, thus indicating that the $\mbox{\boldmath $d$}$-vector is strongly pinned along the $b$ axis.     
Thus, we consider that the characteristic feature of the low-$H$ SC state is the suppression of the spin-susceptibility in $H \parallel b$ (the presence of a finite $\hat{b}$ component in the $\mbox{\boldmath $d$}$ vector).
If so, this SC state can sustain up to $H_{\rm pin}$
and a new SC state sets in above $H_{\rm pin}$, where the $\bm{d}$ vector is perpendicular to the $b$ axis and  $\Delta K_b=0$.
The value of $H_{\rm pin}$ can be estimated from the balance between the SC condensation energy of the low-$H$ superconductivity ($\mu_0 H_c^2 / 2$) and the Zeeman energy at $H_{\rm pin}$ ($\frac{1}{2} \mu_0  \Delta \chi_b H_{\rm pin}^2$). 
Similar discussion was also done for the ferromagnetic superconductor UCoGe\cite{TadaPRB2016}.
$H_{\rm pin}$ is estimated to be $\sim 13$ T from the decrease in spin susceptibility along the $b$ axis $\Delta \chi_b$ derived from the relation  $\Delta \chi_b = \Delta K_{{\rm spin}, b} /A_{{\rm hf}, b}$ with $\Delta K_{{\rm spin}, b} \sim 0.060$\%.
Here, $A_{{\rm hf}, b}$ is a hyperfine coupling constant along the $b$ axis, which was reported as $A_{{\rm hf}, b} = 5.18$ (T /$\mu_B$) from the $K$ - $\chi$ plot in the normal state\cite{TokunagaJPSJ2019}. 
Notably, this $H_{\rm pin}$ is close to the critical field of $H_{\rm {c2}}$, at which $T_{\rm c}$ is minimum in $H$ and increases with $H$, as shown in Fig. 1 (B).
The $H$-boosted superconductivity might be interpreted as a different SC state where the $\mbox{\boldmath $d$}$ vector can rotate away from the $b$ axis.

Recently, Ishizuka and Yanase calculated the maximum magnitude of intrasublattice $\mbox{\boldmath $d$}$-vector components in the whole momentum space. They showed that the predominant $\mbox{\boldmath $d$}$-vector components for the $B_{3u}$ and $A_u$ are $k_b\hat{c}$ and $k_b\hat{b}$, respectively, and these two states are almost degenerate\cite{IshizukaArXiv2020}.
Thus, one interesting scenario to interpret the anomalous $H_{\rm {c2}}$ behavior in $H \parallel b$ is that the low-$H$ (high-$H$) superconductivity corresponds to the $A_u$ ($B_{3u}$) state, and the $\mbox{\boldmath $d$}$-vector rotation occurs at approximately 13 T.  
To verify this case, Knight shift measurement above 13 T in $H \parallel b$ is also crucial, and we expect the $K_b$ decrease to become smaller above 13 T and the $K_b$ to remain  unchanged in the high-$H$ SC state. 

In conclusion, we performed $^{125}$Te-NMR on a single crystal of $^{125}$Te-enriched UTe$_2$ and measured the NMR Knight shift when $H \parallel b$ and $H \parallel c$ below $T_{\rm c}$.
A slight decrease was observed in both $K_b$ and $K_c$ at low $H$; however, $K_c$ remained unchanged at 5.5 T between the SC and normal state.
The latter further supports the spin-triplet scenario, and the former indicates the finite components of $\hat{b}$ and $\hat{c}$ in the SC $\mbox{\boldmath $d$}$ vector.
From the detailed $H$ dependence, although the $\hat{c}$ component is gradually suppressed when $H \parallel c$, the $\hat{b}$ component is finite at least up to 6.5 T, indicating that the $\mbox{\boldmath $d$}$ vector is pinned along the $b$ axis. 
From the estimation of the pinning field $H_{\rm pin}$, we suggest that the SC character would be different between the low-$H$ and high-$H$ SC phases, particularly about the dominant $\mbox{\boldmath $d$}$-vector direction.   
Further experiments to identify the spin state in the high-$H$ SC state, and to clarify the origin of the shoulder peaks in $H \parallel b$ are needed, and now in progress.

\begin{acknowledgments}
The authors would like to thank M. Manago, J. Ishizuka, Y. Yanase, Y. Maeno, S. Yonezawa, and J-P. Brison, G. Knebel, and J. Flouquet for valuable discussions. This work was supported by the Kyoto University LTM Center, Grants-in-Aid for Scientific Research (Grant Nos. JP15H05745, JP17K14339, JP19K03726, JP16KK0106, JP19K14657, JP19H04696, JP19H00646, and JP20H00130) and Grant-in-Aid for JSPS Research Fellows (Grant No. JP20J11939) from JSPS.
\end{acknowledgments}

\bibliography{UTe2}% Produces the bibliography via BibTeX.
\newpage

{\bf \large Supplemental Material}
\section{The comparison of the $^{125}$T\MakeLowercase{e}-NMR spectrum with the previous report}
\begin{figure}[H]
%\vspace*{-10pt}
\begin{center}
\includegraphics[width=6cm,clip]{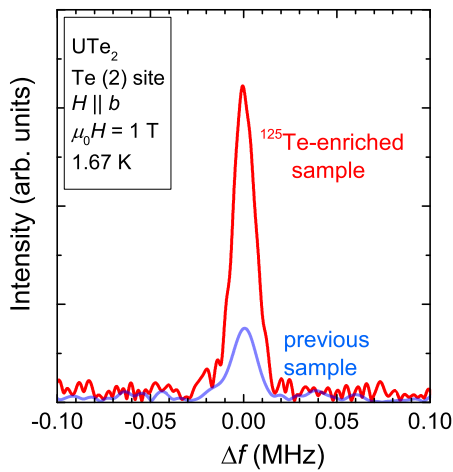}
\end{center}
\caption{\label{FigS.1}Normal-state $^{125}$Te-NMR spectra of the $^{125}$Te-enriched sample used in this study and of a natural-Te sample used on the previous measurement\cite{NakamineJPSJ2019}.
The $\Delta f$ for horizontal axis is the difference in NMR frequency from the peak position($\Delta f = f - f_{\rm peak}$).}

\end{figure}

Figure \ref{FigS.1} shows normal-state $^{125}$Te-NMR spectra of a $^{125}$Te-enriched sample used on this measurement and of a natural-Te sample used on the previous measurement\cite{NakamineJPSJ2019}.

The NMR-signal intensity of the enriched sample is much stronger than that of the previous sample, although the sample volume of two samples is almost the same, 
In addition, the linewidth of the two samples is almost the same, ensuring that the sample quality of the enriched one is unchanged with the natural one.
%%%%%%%%%%%%%%%%%%%%%%%%%%%% Figure S1 %%%%%%%%%%%%%%%%%%%%%%%%%%%%%%%%%%%%%
%%%%%%%%%%%%%%%%%%%%%%%%%%%%%%%%%%%%%%%%%%%%%%%%%%%%%%%%%%%%%%%%%%%%%%%%%%%
\section{Alignment of the magnetic field }
\begin{figure}[H]
%\vspace*{-10pt}
\begin{center}
\includegraphics[width=8.7cm,clip]{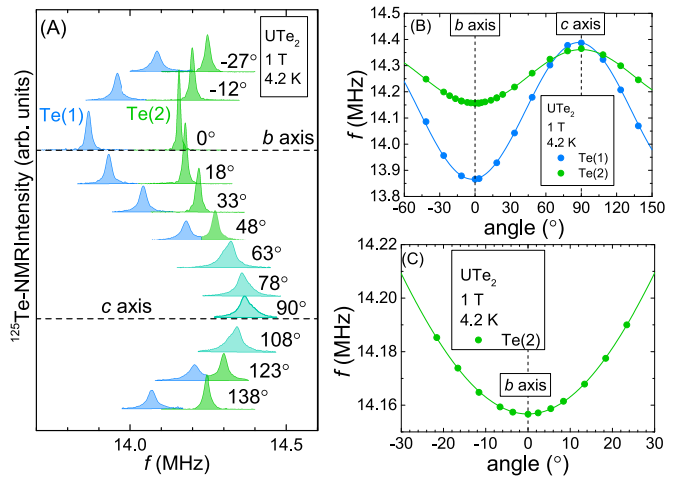}
\end{center}
\caption{\label{FigS.2}(A) The angular dependence of the $^{125}$Te-NMR spectra of both the Te(1) and Te(2) sites at 4.2 K.
The magnetic field is 1 T in the $bc$ plane.
(B) The angular dependence of the resonance peaks for both the Te(1) and Te(2) sites.
(C) The zoomed view of the angular dependence of the resonance frequency of the Te(2) site around the $b$ axis.}

\end{figure}

In order to apply magnetic field exactly parallel to the $b$ or $c$ axis, we used the split superconducting (SC) magnet generating a horizontal field combined with a single-axis rotator with the $a$-axis being the rotation axis.
Figure \ref{FigS.2} shows the angular dependence of $^{125}$Te-NMR spectra (A) and the resonance peaks (B) of both the Te(1) and Te(2) sites at 4.2 K under the field of 1 T in the $bc$ plane.
The obtained angular dependence is consistent with the previous reports\cite{TokunagaJPSJ2019, NakamineJPSJ2019}.
As well as the previous study\cite{NakamineJPSJ2019}, we measured the spectra near the $b$ axis with a small angle step and plotted the resonance peak as shown in Fig.~\ref{FigS.2} (C), and aligned the field exactly parallel to the $b$ axis.
%%%%%%%%%%%%%%%%%%%%%%%%%%%% Figure S2 %%%%%%%%%%%%%%%%%%%%%%%%%%%%%%%%%%%%%
%%%%%%%%%%%%%%%%%%%%%%%%%%%%%%%%%%%%%%%%%%%%%%%%%%%%%%%%%%%%%%%%%%%%%%%%%%%
\section{Evaluation of the Knight shift and linewidth of NMR spectrum in $\bm{H \parallel c}$}
As shown in Fig.~S2, the Te(1) and Te(2) signals were distinct when $H \parallel b$ but overlapped when $H \parallel c$.
It is difficult to distinguish between the two signals because the line width is broad in $H \parallel c$.
Therefore, we determined the Knight shift from the spectral-peak frequency because the contribution of the Te(2) site was dominant even in the broad spectrum when $H \parallel c$.
In addition, we defined the linewidth as the left-side half width at half maximum (LHWHM) as shown in the inset of Fig.~\ref{FigS.3} to avoid the contribution of the Te(1) signal.
The broken curves in Figs. 1(D), 2(B), 2(C), and the inset of Fig. S3 indicating the Te(1) and Te(2) sites are guide to eye and are depicted as follows. 
As seen in $H \parallel b$, the signal intensity ratio of Te(1) and Te(2) is not 1:1 by some reasons such as difference of nuclear spin-spin relaxation rate, and of pulse conditions.
Thus, the broken curves were determined by double Lorentzian fitting with the fixed signal area ratio [Te(1)/Te(2)~0.56], referring to the spectrum reported by Tokunaga {\it et al.}\cite{TokunagaJPSJ2019}
\section{temperature dependence of the linewidth of the NMR spectrum} %% No sections necessary for express letters, letters and short notes
In general, the linewidth of the NMR spectrum becomes broader in the SC state than that in the normal state, by the presence of the SC diamagnetic shielding and the distribution of the spin susceptibility.
Therefore, the broadening of the linewidth below $T_{\rm c}$ is one of the reliable confirmation that the NMR spectrum in the SC state is measured.
%%%%%%%%%%%%%%%%%%%%%%%%%%%% Figure S3 %%%%%%%%%%%%%%%%%%%%%%%%%%%%%%%%%%%%%
\begin{figure}[H]
%\vspace*{-10pt}
\begin{center}
\includegraphics[width=5.5cm,clip]{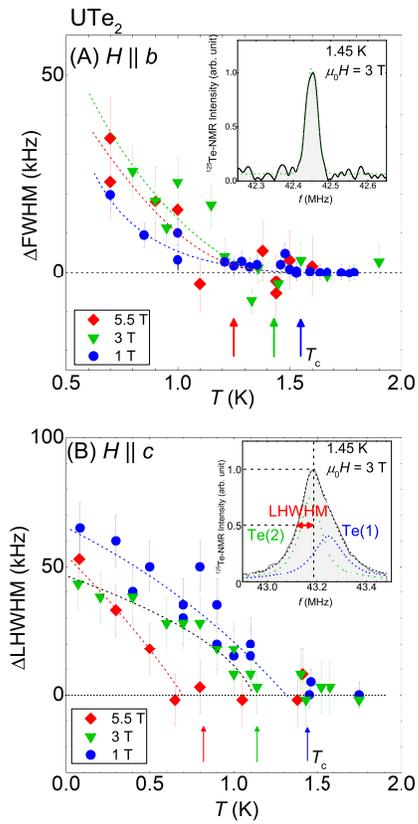}
\end{center}
\caption{\label{FigS.3}The temperature dependence of the increase of FWHM in the SC state ($\Delta$FWHM ) when $H \parallel b$ (A) and of the increase of left-side HWHM in the SC state when $H \parallel c$ (B).
$\Delta$FWHM is defined as $\Delta$FWHM $\equiv$ FWHM($T$) - FWHM in the normal state.
Dashed lines are added as guide to eye and arrows show superconducting transition temperature obtained from the AC susceptibility measurements. 
(inset) The typical NMR spectrum in $H \parallel c$ to show the definition of left-side HWHM. 
Dotted lines represents the result of fitting by the two-Lorentzian function.}
\end{figure}
%%%%%%%%%%%%%%%%%%%%%%%%%%%%%%%%%%%%%%%%%%%%%%%%%%%%%%%%%%%%%%%%%%%%%%%%%%%
Figure~\ref{FigS.3} shows the behavior of the linewidth increase of the NMR spectrum for $H\parallel b$ (A) and $H \parallel c$ (B).
The arrows in the figure show the $T_{\rm c}$ determined with the AC susceptibility measurement.
The full width at the half maximum (FWHM) in $H\parallel b$ is obtained by the Gaussian fitting of the whole spectrum.
The FWHMs below $\sim 0.8$ K are not plotted because a shoulder peak appears at the lower-frequency side (i.e. at the lower Knight shift) than the original peak frequency below 0.9 K, as mentioned in the main text.
The spectrum in $H \parallel b$ is broadened from just below $T_{\rm c}$.
When $H \parallel c$, we plotted LHWHM as mentioned above.
The spectra in $H\parallel c$ is also broadened from just below $T_{\rm c}$, as in the case when $H \parallel b$.
The broadening of the spectrum from just below $T_{\rm c}$ when $H\parallel b$ and $H\parallel c$ ensures that the NMR spectrum in the SC state is certainly measured.

\end{document}